\documentclass[12pt]{article}
\usepackage{geometry,amsmath,amssymb,graphicx}

\newcommand{\mathe}{\mathrm{e}}
\newcommand{\tmem}[1]{{\em #1\/}}
\newcommand{\tmop}[1]{\ensuremath{\operatorname{#1}}}
\newcommand{\tmtextbf}[1]{{\bfseries{#1}}}
\newcommand{\tmtextit}[1]{{\itshape{#1}}}

\begin{document}

\title{Performance of a worm algorithm in $\phi^4$ theory at finite quartic
coupling}
\author{Tomasz Korzec, Ingmar Vierhaus and Ulli Wolff\thanks{
e-mail: uwolff@physik.hu-berlin.de} \\
Institut f\"ur Physik, Humboldt Universit\"at\\ 
Newtonstr. 15 \\ 
12489 Berlin, Germany
}
\date{}
\maketitle

\begin{abstract}
 Worm algorithms have been very successful with the simulation of sigma
  models with fixed length spins which result from scalar field theories in
  the limit of infinite quartic coupling $\lambda$. Here we investigate closer
  their algorithmic efficiency at finite and even vanishing $\lambda$ 
for the one component model in
  dimensions $D = 2, 3, 4$.
\end{abstract}

\begin{flushright} HU-EP-11/03 \end{flushright}
\thispagestyle{empty}
\newpage

\section{Introduction}

In reference {\cite{prokofev2001wacci}} it was shown how spin systems of the
Ising and XY model type can be Monte Carlo simulated with greatly reduced or
even eliminated critical slowing down by what the authors call worm
algorithms. They can be seen as stochastically sampling strong coupling or
hopping parameter expansion graphs of these models to in principle arbitrary
order instead of the original spin or field configurations. The method has
been generalized in several directions, including sigma models of the O($N$)
type, and a summary with further references can be found in
{\cite{Wolff:2010zu}}. All these systems can be regarded as
linear{\footnote{Spins live in the linear space $\mathbb{R}^N$ but, of
course, are still nonlinearly coupled in general.}} sigma or $\phi^4$ models,
where in the limit of infinite self coupling $\lambda$ the length fluctuations
of the spins are frozen to $N - 1$ dimensional spheres. In this paper it is
our goal to systematically investigate for $N = 1$ the dynamical behavior of
worm algorithms away from the Ising limit at {\tmem{finite}} $\lambda$ and
even in the Gaussian limit $\lambda = 0$ in Euclidean space dimensions $D = 2,
3, 4$. Results for the Ising limit $\lambda = \infty$ may be inspected for
comparison in {\cite{Deng:2007jq}} and {\cite{Wolff:2008km}}. The paper is
organized as follows. In Section 2 we discuss the reformulation of $\phi^4$
theory and of a few observables by the all-order strong coupling
contributions. This is followed by Section 3 detailing our update algorithm and
presenting our results on critical slowing down. In Section 4 we summarize our
conclusions and in Appendix~\ref{app} we compile data for $\lambda = 1 / 2$ that can be
useful as a reference for future studies. Our report here is a summary of a
diploma thesis available under {\cite{VierhausD}} with a lot more details.

\section{(Re-)formulation of the model}

\subsection{Partition function}

We use lattice units ($a = 1$) and start from the standard lattice formulation
of scalar field theory with the partition function
\begin{equation}
  Z_0 = \int \left[ \prod_z d \mu_{\lambda} (\phi (z)) \right] \mathe^{\beta
  \sum_{l = \langle x y \rangle} \phi (x) \phi (y)}
\end{equation}
where the measure at each site $z$ is given by
\begin{equation}
  \int d \mu_{\lambda} (\phi) f (\phi) = \frac{\int_{- \infty}^{\infty} d \phi
  \mathe^{- \phi^2 - \lambda (\phi^2 - 1)^2} f (\phi)}{\int_{-
  \infty}^{\infty} d \phi \mathe^{- \phi^2 - \lambda (\phi^2 - 1)^2}}
  \label{measure}
\end{equation}
and thus includes the coupling $\lambda \geqslant 0$. Our hypercubic lattice
is wrapped on a $D$ dimensional torus with $L$ sites in each direction, and
the sum is over links $l$ corresponding to nearest neighbor pairs of sites
$\langle x y \rangle$. Below we shall need the moments
\begin{equation}
  c_{\lambda} (n) = \int d \mu_{\lambda} (\phi) \phi^n \label{moments}.
\end{equation}
In particular for the $\lambda$ values simulated in this study the
nonvanishing even moments are
\begin{eqnarray}
  c_0 (2 n) & = & \Gamma (n + 1 / 2) / \Gamma (1 / 2) = 2^{- n} (2 n - 1) !!,
  \\
  c_{1 / 2} (2 n) & = & 2^{n / 2} \Gamma (n / 2 + 1 / 4) / \Gamma (1 / 4), \\
  c_{\infty} (2 n) & = & 1, 
\end{eqnarray}
while odd moments vanish due to Z(2) symmetry.

The fundamental two point correlation function is regarded as a ratio
\begin{equation}
  G (u - v) = \langle \phi (u) \phi (v) \rangle = \frac{Z (u, v)}{Z_0}
\end{equation}
with the `partition function with insertions' as numerator
\begin{equation}
  Z (u, v) = \int \left[ \prod_z d \mu_{\lambda} (\phi (z)) \right]
  \mathe^{\beta \sum_{l = \langle x y \rangle} \phi (x) \phi (y)} \phi (u)
  \phi (v) .
\end{equation}
By expanding in $\beta$ the Boltzmann factor on each link and then integrating
over $\phi$ independently on each site using (\ref{moments}) the same quantity
is given by
\begin{equation}
  Z (u, v) = \sum_k \left[ \prod_l \frac{\beta^{k (l)}}{k (l) !} \right]
  \prod_x c_{\lambda} (\partial k (x) + \delta_{x, u} + \delta_{x, v}) .
\end{equation}
Here the sum is over a link field that is independently summed over $k (l) =
0, 1, \ldots, \infty$ and the divergence
\begin{equation}
  \partial k (x) = \sum_{l, \partial l \ni x} k (l)
\end{equation}
counts the sum of $k (l)$ over all links surrounding a site $x$. We have a
well defined non-negative weight for any $k$ configuration. Nonzero
contributions arise if $\partial k$ is even everywhere except the sites $u$,
$v$ if they do not coincide, where $\partial k$ must be odd. Finally we
introduce the ensemble{\footnote{A weight depending on the locations $u, v$
could be included as in {\cite{Wolff:2008km}}, but we have not yet explored
this generalization.}}
\begin{equation}
  \mathcal{Z}= \sum_{u, v} Z (u, v) = \sum_{k, u, v} \left[ \prod_l
  \frac{\beta^{k (l)}}{k (l) !} \right] \prod_x c_{\lambda} (\partial k (x) +
  \delta_{x, u} + \delta_{x, v}) \label{Zens}
\end{equation}
which will be simulated by the worm algorithm. In this way expectation values
\begin{equation}
  \langle \langle \mathcal{O}[k ; u, v] \rangle \rangle =
  \frac{1}{\mathcal{Z}} \sum_{k, u, v} \left[ \prod_l \frac{\beta^{k (l)}}{k
  (l) !} \right] \left[ \prod_x c_{\lambda} (\partial k (x) + \delta_{x, u} +
  \delta_{x, v}) \right] \mathcal{O}[k ; u, v]
\end{equation}
become accessible to Monte Carlo estimation.

\subsection{Observables}

The fundamental two point function is now given by
\begin{equation}
  G (x) = \langle \phi (x) \phi (0) \rangle = \frac{\langle \langle \delta_{x,
  u - v} \rangle \rangle}{\left\langle \left\langle \delta_{u, v} r_{\lambda}
  (\partial k (u)) \right\rangle \right\rangle} \label{correst}
\end{equation}
where we have introduced
\begin{equation}
  r_{\lambda} (2 n) = \frac{c_{\lambda} (2 n)}{c_{\lambda} (2 n + 2)},
\end{equation}
which is only required and defined for even arguments. For universal
quantities only ratios of two point functions at differing separations are of
interest and then the denominator is not required. It is however related to
the susceptibility
\begin{equation}
  \chi = \sum_x G (x), \hspace{1em} \chi^{- 1} = \left\langle \left\langle
  \delta_{u, v} r_{\lambda} (\partial k (u)) \right\rangle \right\rangle .
  \label{chidef}
\end{equation}
The second moment (renormalized) mass $m$ can be defined and then measured by
\begin{equation}
  \frac{\sum_x \cos (2 \pi x_{\mu} / L) G (x)}{\chi} = \frac{m^2}{m^2 + 4
  \sin^2 (\pi / L)} = \langle \langle f_m (u - v) \rangle \rangle \Rightarrow
  m \label{mdef}
\end{equation}
with
\begin{equation}
  f_m (x) = \frac{1}{D} \sum_{\mu} \cos (2 \pi x_{\mu} / L) .
\end{equation}
Note that in our symmetric setup each of the $D$ directions contributes
equally and we average for better statistics.

An estimator for the energy (without the measure part) is given by
\begin{equation}
  E = \frac{1}{D L^D}  \frac{\partial}{\partial \beta} \ln Z_0 =
  \frac{1}{\beta D L^D}  \frac{\langle \langle \delta_{u, v} r_{\lambda}
  (\partial k (u)) \sum_l k (l) \rangle \rangle}{\left\langle \left\langle
  \delta_{u, v} r_{\lambda} (\partial k (u)) \right\rangle \right\rangle} .
  \label{Edef}
\end{equation}
A second estimator for $E$ is given by the nearest neighbor correlation using
(\ref{correst}). As in the Ising limit {\cite{Wolff:2008km}} the error is much
larger in this case.

\section{Algorithm and dynamical results}

In the first Subsection we describe details of the worm algorithm implemented
by us. The code was written in C and has run on standard dual quad-core PCs.
The data reported below correspond to a few core-months total runtime
dominated by the largest lattices of sizes $128^3$ and $32^4$.

\subsection{Specification of the update scheme}

The update algorithm used by us is a generalization of the one described in
{\cite{Wolff:2008km}}. We define two types of micro-steps that each obey
detailed balance with respect to the ensemble (\ref{Zens}):
\begin{itemize}
  \item $I$: With equal probability we pick one of the $2 D$ neighbors $u'$ of
  the present configuration's $u$ and denote by $l$ the link in between. Again
  with equal probability we propose one of the two moves $k (l) \rightarrow k
  (l) \pm 1$ accompanied in either case by the change $u \rightarrow u'$. The
  proposal is accepted with the respective Metropolis probabilities $\min (1,
  q_{I \pm})$ where we take
  \begin{equation}
    q_{I +} = \frac{\beta}{k (l) + 1} r_{\lambda}^{- 1} (\partial k (u') +
    \delta_{u', v})
  \end{equation}
  and
  \begin{equation}
    q_{I -} = \frac{k (l)}{\beta} r_{\lambda} (\partial k (u) + \delta_{u, v}
    - 1) .
  \end{equation}
  \item $II$: We only act if $u = v$ holds, and even then only with
  probability $p_0 = 1 / 2$. Then a randomly chosen new location $u' = v'$ (at
  unchanged $k$) is proposed and accepted with probability $\min (1, q_{I
  I})$,
  \begin{equation}
    q_{II} = \frac{r_{\lambda} (\partial k (u))}{r_{\lambda} (\partial k
    (u'))} .
  \end{equation}
\end{itemize}
Swapping the r\^oles of $u$ and $v$, step $I$ can also be applied to move $v$
and we call these possibilities now $I_u$ and $I_v$. As an iteration we define
the $L^D / 2$ fold repetition of the sequence $I_u\,II\,I_v\,II$ which is
similar to a sweep of a standard local algorithm as far as CPU work is
concerned.

\subsection{Monte Carlo dynamics in the Gaussian limit}

For the Gaussian case $\lambda = 0$ all our observables can be computed
exactly by straight forward Fourier expansion. We therefore do not list any
mean values in this case but we have monitored that our results were correct
within errors. The dynamics of the `worm' algorithm remains of interest at
$\lambda = 0$ and is presumably representative for other small bare couplings.
To investigate critical slowing down we keep constant the extension in
physical units $m L$ while we increase $L$. In the free case this leads to
choosing
\begin{equation}
  \beta^{- 1} = D + \frac{m^2}{2} .
\end{equation}
\begin{figure}[htb]
\begin{center}
  \resizebox{0.5\textwidth}{!}{\includegraphics{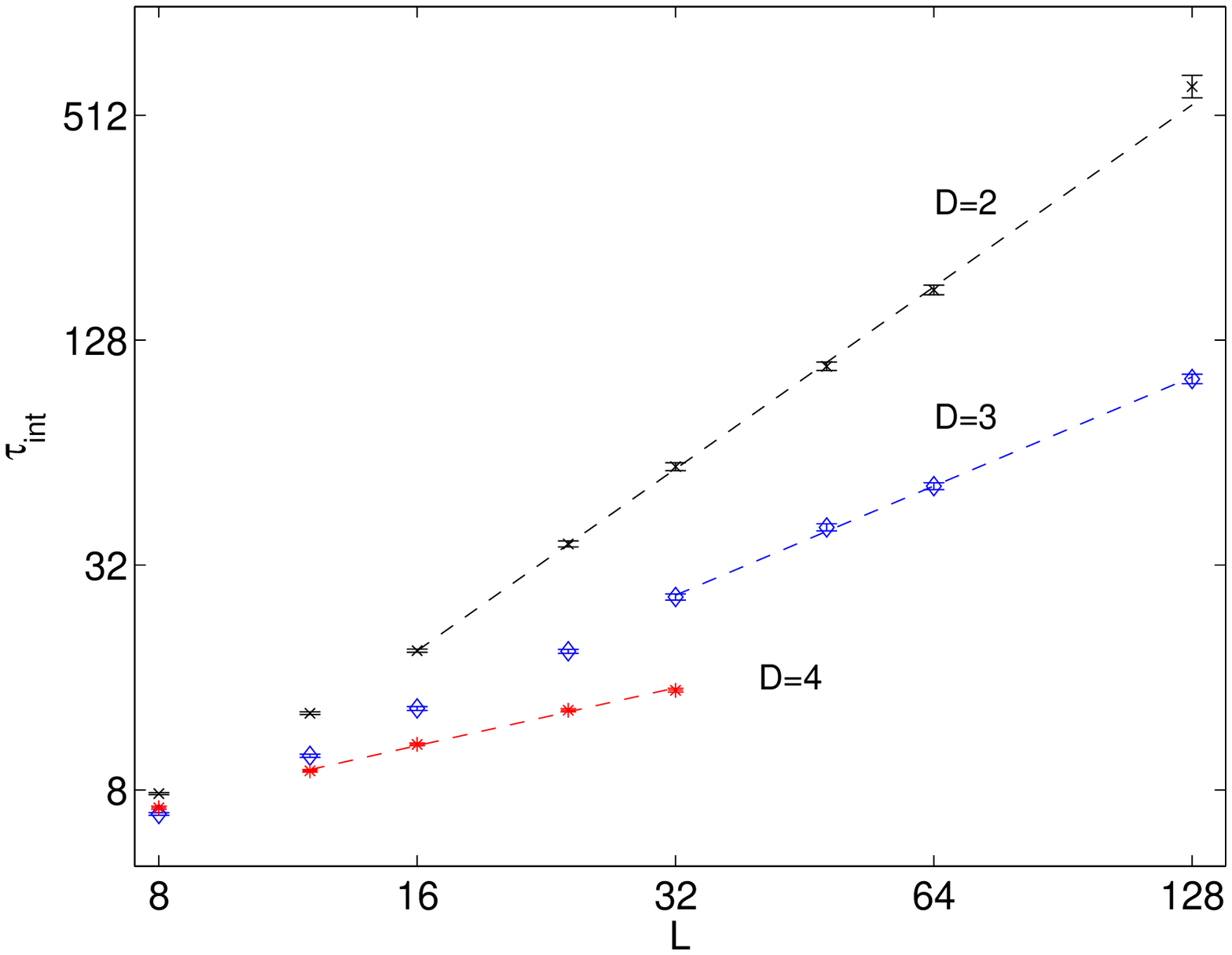}}\resizebox{0.5\textwidth}{!}{\includegraphics{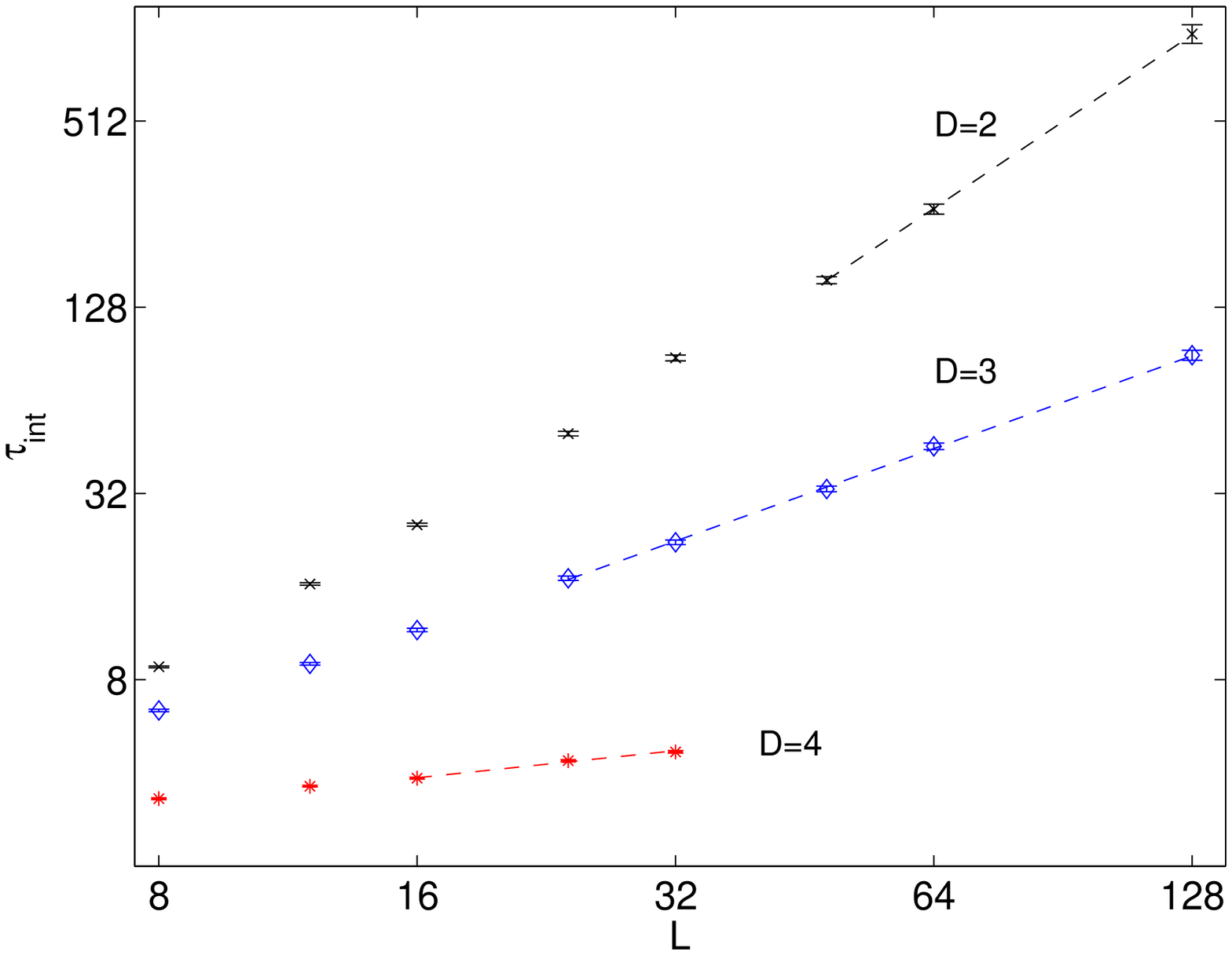}}
  \caption{Integrated autocorrelation times for the energy $E$ (left panel)
  and the mass $m$ (right panel) for the Gaussian model in dimensions $D = 2,
  3, 4$ with $m L = 4$. The dashed lines are fits of the form
  $\tau_{\tmop{int}} \propto L^z$, $z$-values are in the text. \label{fig1}}
\end{center}
\end{figure}

All autocorrelation times were determined as discussed in
{\cite{Wolff:2003sm}} and are given in units of `iterations' defined above.
Measurements are taken and pre-averaged during each iteration and then stored
for off-line analysis. In Fig.~\ref{fig1} we see log-log plots of integrated
autocorrelation times for the observables $E$ and $m$. Note that the estimator
(\ref{Edef}) refers to a ratio of primary Monte Carlo estimated mean values
and we refer to {\cite{Wolff:2003sm}} for the definition of
$\tau_{\tmop{int}}$ for such derived quantities. The dashed lines are fits
with dynamical exponents $z$. From top to bottom we have determined $z = 1.62
(2), 0.97 (3), 0.51 (2)$ (left plot) and $z = 1.86 (7), 0.99 (2), 0.29 (2)$
(right plot). The quoted errors are purely statistical. We consider these fits
which have have acceptable $\chi^2$ over the range shown as mere
parameterizations of our data in the range where most simulations work. We
have not embarked on the difficult assessment of systematic errors with regard
to truly asymptotic dynamical behavior. In summary we see here distinct
critical slowing down, weaker than for standard local methods but inferior to
worm (and also cluster) simulations in the Ising limit. There is a pronounced
tendency of decorrelation improving with increasing dimension. The
susceptibility $\chi$ was also investigated and behaves similarly to $m$. The
ultraviolet quantity $G (0)$ measured via (\ref{correst}) on the other hand
typically exhibits shorter autocorrelations. We also have a large set of data
in smaller volumes $m L = 1$. Autocorrelations are larger for this more
critical series but the overall qualitative behavior is quite similar. Further
details can be found in {\cite{VierhausD}}.

\subsection{Monte Carlo dynamics at intermediate coupling}

\begin{figure}[htb]
\begin{center}
  \resizebox{0.5\textwidth}{!}{\includegraphics{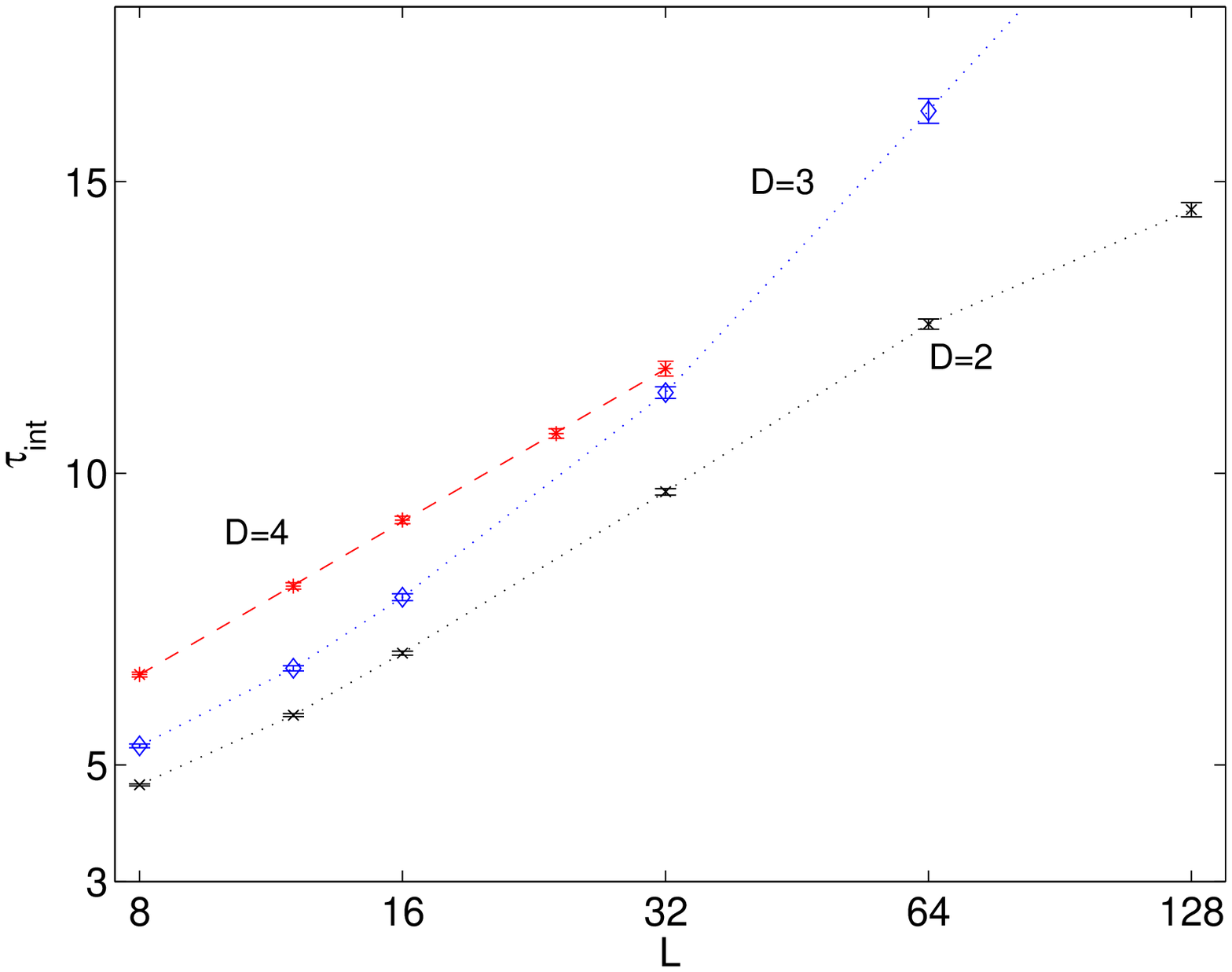}}\resizebox{0.5\textwidth}{!}{\includegraphics{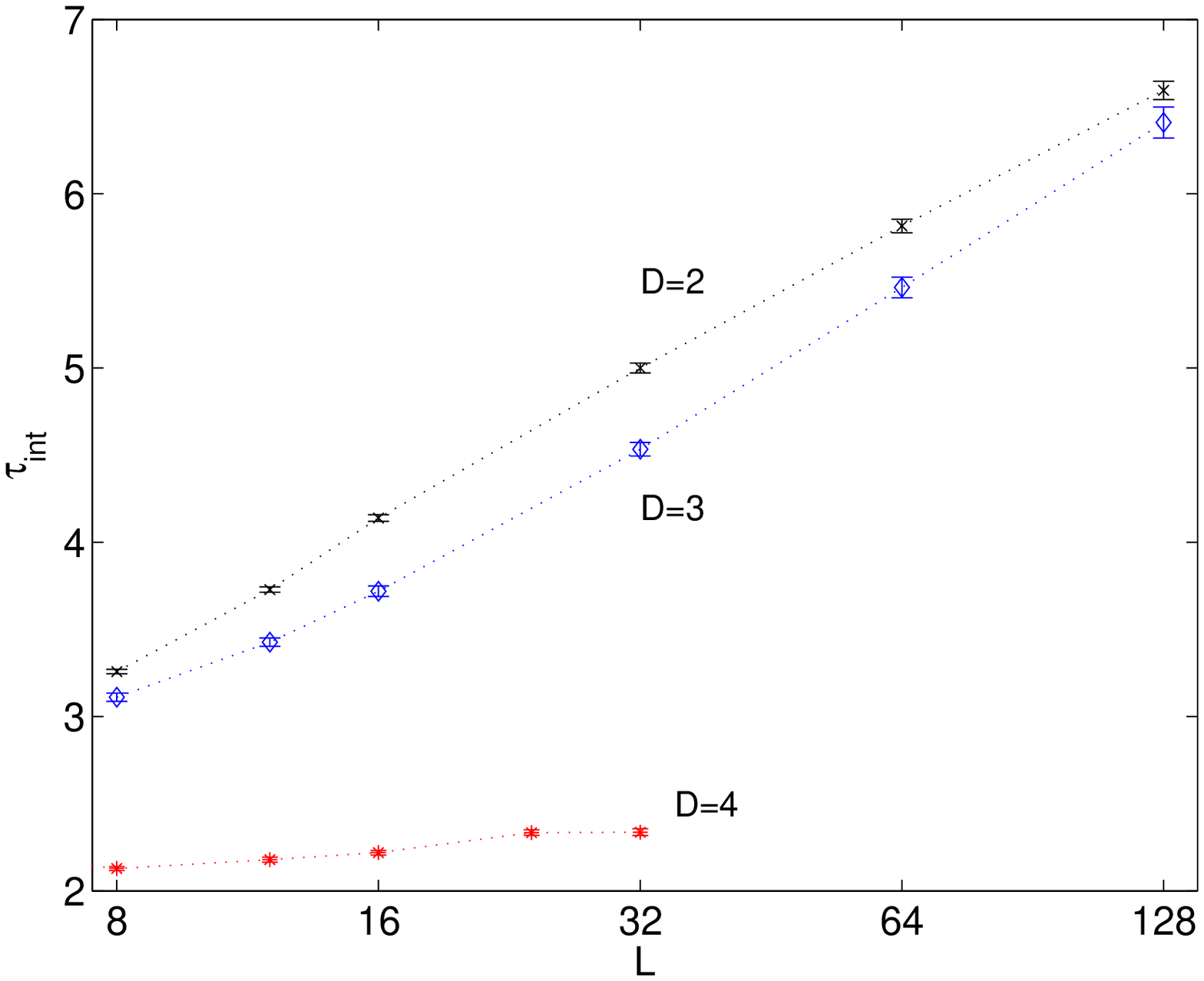}}
  \caption{Integrated autocorrelation times for energy $E$ (left panel) and
  mass $m$ (right panel) for $\lambda = 1 / 2$ in dimensions $D = 2, 3, 4$
  with $m L = 4$. The dashed line is a fit for $\tau_{\tmop{int}}$ linear in
  $\ln L$, while the dotted lines just connect data points to guide the eye.
  \label{fig2}}
\end{center}
\end{figure}

We now turn to simulations at $\lambda = 1 / 2$ which are otherwise organized
similarly to those of the previous subsection. As the values of our
observables are non-trivial we collect them in Appendix~\ref{app} for this
case. The dynamical results are given in Fig.~\ref{fig2}.

We have chosen semilogarithmic plots as the most natural appearing common
format here. Two term fits linear in $\ln L$ do however not quite achieve good
$\chi^2$ values except for $E$ in $D = 4$ (dashed line). What is clearly
visible is that the worm algorithm works much better in this interacting case
then in the Gaussian limit. Although precise fits require more than the first
power of $\ln L$ (or larger $L$ to assume the asymptotic form) the general
behavior looks close to an only logarithmic growth of autocorrelations. The
tendency with $D$ is mild and the ordering differs for the two observables
displayed. Also at this coupling we have looked at further observables and
have also simulated $m L = 1$ with qualitatively similar outcomes.

\section{Conclusions}

We have found that a worm algorithm is applicable to $\phi^4$ theory all the
way to the Gaussian limit in dimensions between two and four. Although still
exhibiting less critical slowing down than standard local methods, the free
limit $\lambda = 0$ is the most difficult case for the worm algorithm. For the
intermediate coupling $\lambda = 1 / 2$ the situation is already similar to
the infinite coupling limit with regard to decorrelation. While at zero
coupling the efficiency shows pronounced improvement with growing dimension,
this dependence is much weaker at $\lambda = 1 / 2$. In {\cite{Wolff:2009ke}}
a very dramatic improvement was realized in the Ising limit by the
construction of a special estimator for the connected four point function
without the need to perform numerical cancellations. We have so-far not
succeeded in generalizing this construction to finite $\lambda$ in a similarly
efficient way. Some such efforts are however reported in {\cite{VierhausD}}.

{\noindent}\tmtextbf{Acknowledgments}. We thank Martin Hasenbusch for
discussions. Financial support of the DFG via SFB transregio 9 is
acknowledged.

\appendix\section{Results of simulations at $\lambda=1/2$ }
\label{app}
In this appendix we report on some mean values of observables defined in
(\ref{mdef}), (\ref{Edef}), (\ref{chidef}) together with $\beta$ values that
have been determined by tuning to $m L = 4$ within errors. The data are
summarized in tables \ref{tab2}, \ref{tab3}, \ref{tab4}. The last column holds
the numbers of iterations performed. We note that the mass $m$ has a
practically constant and mostly sub per mille error at -- or rescaled to -- a
constant iteration number. This confirms the very mild or even absent slowing
down.

\begin{table}[htb]
\begin{center}
  \begin{tabular}{|c|c|c|c|r@{.}l|c|}
    \hline
    $L$ & $\beta$ & $m L$ & $E$ &  \multicolumn{2}{c|}{$\chi$} & its/$10^6$ \\
    \hline
    8 & 0.576950 & 4.0036(20) & 0.26184(13) & 6&566(5) & 24\\
    \hline
    12 & 0.615670 & 3.9982(22) & 0.32214(13) & 13&381(11) & 24\\
    \hline
    16 & 0.634350 & 4.0008(23) & 0.36241(12) & 22&274(20) & 24\\
    \hline
    22 & 0.649270 & 4.0028(25) & 0.40381(11) & 39&334(38) & 24\\
    \hline
    32 & 0.661390 & 3.9991(26) & 0.44622(9)\phantom{1} & 76&855(80) & 24\\
    \hline
    64 & 0.674330 & 4.0067(29) & 0.50603(7)\phantom{1} & 262&87(30) & 24\\
    \hline
    128 & 0.680679 & 3.9936(35) & 0.54516(5)\phantom{1} & 898&3(1.2) & 20\\
    \hline
  \end{tabular}
  \caption{Results at $D = 2$.\label{tab2}}
\end{center}
\end{table}

\begin{table}[htb]
\begin{center}
  \begin{tabular}{|c|c|c|c|r@{.}l|c|}
    \hline
    $L$ & $\beta$ & $m L$ & $E$ &  \multicolumn{2}{c|}{$\chi$} & its/$10^6$ \\
    \hline
    8 & 0.370240 & 3.9994(26) & 0.14561(7) & 10&610(12) & 8\\
    \hline
    12 & 0.383030 & 3.9984(27) & 0.16235(5) & 22&915(28) & 8\\
    \hline
    16 & 0.388310 & 3.9957(28) & 0.17178(4) & 39&963(50) & 8\\
    \hline
    22 & 0.391920 & 4.0007(29) & 0.17989(3) & 74&06(10) & 8\\
    \hline
    32 & 0.394390 & 3.9999(30) & 0.18692(2) & 154&08(21) & 8\\
    \hline
    64 & 0.396400 & 3.9998(32) & 0.19457(1) & 600&02(90) & 8\\
    \hline
    128 & 0.397067 & 4.0059(49) & 0.19816(1) & 2336&9(5.3) & 4\\
    \hline
  \end{tabular}
  \caption{Results at $D = 3$.\label{tab3}}
\end{center}
\end{table}

\begin{table}[htb]
\begin{center}
  \begin{tabular}{|c|c|c|c|r@{.}l|c|}
    \hline
    $L$ & $\beta$ & $m L$ & $E$ &  \multicolumn{2}{c|}{$\chi$} & its/$10^6$\\
    \hline
    4 & 0.245490 & 4.0021(20) & 0.079119(52) & 4&0522(34) & 8\\
    \hline
    8 & 0.271670 & 4.0012(17) & 0.092157(16) & 14&624(14) & 8\\
    \hline
    12 & 0.277630 & 3.9977(15) & 0.097502(8)\phantom{1} & 32&183(23) & 8\\
    \hline
    16 & 0.279870 & 4.0003(15) & 0.099984(5)\phantom{1} & 56&669(39) & 8\\
    \hline
    24 & 0.281560 & 3.9972(14) & 0.102197(3)\phantom{1} & 126&869(83) & 8\\
    \hline
    32 & 0.282173 & 4.0007(19) & 0.103132(4)\phantom{1} & 224&41(20) & 4\\
    \hline
  \end{tabular}
  \caption{Results at $D = 4$.\label{tab4}}
\end{center}
\end{table}

\end{document}